\documentclass[11pt,a4paper]{article}
\pdfoutput=1
\usepackage{amsmath,amssymb}
\usepackage{epsfig,euscript,array}
\usepackage{amssymb}
\usepackage{graphics}
\usepackage{graphicx}

\def\a{\alpha}
\def\b{\beta}
\def\g{\gamma}

\def\D{\Delta}
\def\d{\delta}

\def\beq{\begin{equation}}
\def\eeq{\end{equation}}
\def\beqn{\begin{eqnarray}}
\def\eeqn{\end{eqnarray}}
\def\ba{\begin{eqnarray}}
\def\ea{\end{eqnarray}}

\def\m{{\tt -}}

\def\l{\langle}

\def\xprim2bar{\overline{x}^{\prime\prime}}

\def\beq{\begin{equation}}
\def\eeq{\end{equation}}

\newcommand{\beqa}{\begin{eqnarray}}
\newcommand{\eeqa}{\end{eqnarray}}

\let\a=\alpha   \let\b=\beta   \let\g=\gamma   \let\d=\delta
 \let\z=\zeta        
      \let\l=\lambda  \let\m=\mu

  \let\D=\Delta

\newcommand{\re}{{\rm e}}

\newcommand{\rD}{{\rm D}}

\newcommand{\oV}{\overline V}

\newcommand{\ov}{\overline v}

\let\a=\alpha   \let\b=\beta   \let\g=\gamma   \let\d=\delta
 \let\z=\zeta        
      \let\l=\lambda  \let\m=\mu

  \let\D=\Delta

\newcommand{\be}{\begin{equation}}
\newcommand{\ee}{\end{equation}}
\newcommand{\bea}{\begin{eqnarray}}
\newcommand{\eea}{\end{eqnarray}}

\newcommand{\eq}[1]{Eq.~(\ref{#1})}

\def\A5{(A_5)_{\rm lat}}
\def\thintablerule{\hrule height0.4pt}
\begin{document}

\vskip 1.5cm
\centerline{\LARGE On the large $N$ limit of $SU(N)$ lattice gauge theories}
\vskip 0.3cm
\centerline{\LARGE  in five dimensions}

\vskip 2 cm
\centerline{\large Nikos Irges and George Koutsoumbas}
\vskip1ex
\vskip.5cm
\centerline{\it Department of Physics}
\centerline{\it National Technical University of Athens}
\centerline{\it Zografou Campus, GR-15780 Athens Greece}
\vskip 1.5 true cm
\thintablerule
\vskip 2.0ex
\leftline{\bf Abstract}
We develop the necessary tools for computing fluctuations around a mean-field background in
the context of $SU(N)$ lattice gauge theories in five dimensions.
In particular, expressions for the scalar observable and the Wilson Loop are given.
As an application, using these observables we compute a certain quantity $k_5$ that
can be viewed as Coulomb's constant in five dimensions.
We show that this quantity becomes independent of $N$ in the large $N$ limit.
Furthermore, the numerical value of $k_5$ we find for $SU(\infty)$ deviates by $17\%$ from its
value predicted by holography.

\vskip 1.0ex\noindent
\vskip 2.0ex
\thintablerule

\vskip-0.2cm
\newpage

\section{Introduction}

In four dimensions and in infinite volume $SU(N)$ gauge interactions are renormalizable and always confining.
Five-dimensional (5d) $SU(N)$ gauge theories are non-renormalizable and trivial at the perturbative limit.
Their weak coupling phase is Coulomb and
they have a first order phase transition at a critical value of the coupling.
Beyond the phase transition there is a confined phase.
The appropriate analytical method to describe these theories near the phase transition
from the side of the Coulomb phase is the Mean-Field (MF) expansion \cite{Review}.
The motivation to consider these theories, beyond the possibility of the existence of a
physical fifth dimension, is that the MF method becomes more accurate as
the dimension of space-time increases. This gives us the opportunity to
develop a trustable analytical probe away from the perturbative point.
Then one can look at regimes where the system is dimensionally reduced and
build effectively four-dimensional (4d) models.
This would not be possible directly in four dimensions as the MF background
vanishes identically in the confined phase.
As we argue here however, there are interesting theoretical issues that can be discussed already in five dimensions,
in particular related to the large $N$ limit of these theories and their possible holographic description.

In \cite{MF11} the five-dimensional $SU(2)$ MF formalism was developed, formulated on a periodic and
anisotropic lattice. We will be using this work's notation and results extensively.
In a first application  \cite{MF12} it was shown that there is a regime on the
phase diagram where the system reduces dimensionally by localizing the gauge interactions
on four-dimensional hyperplanes. Interactions on the hyperplanes are confining and the string tension
as well as some of its corrections, such as the Luscher term, were computed.
The localization of the gauge interactions in this regime was consequently supported also by Monte Carlo
simulations in \cite{FrancMagdAnt}. In the meantime there has been a steadily increasing activity in
lattice simulations of five dimensional gauge theories \cite{5dlatticeTorus}.
In \cite{MF21} the $SU(2)$ MF model with orbifold boundary conditions is constructed,
the goal being a non-perturbative description of Gauge-Higgs Unification.
These (semi)analytical computations complement earlier lattice orbifold Monte Carlo simulations
\cite{orbMC}.

Another application of the MF formalism, not relying on dimensional reduction and one which will
be of our main interest here, was presented in \cite{5dCoulomb} where
a certain quantity, an analogue of Coulomb's constant in five dimensions was
computed. It is derived from the  five-dimensional Coulomb static potential $V_5(r) = {\rm const. - c_2/r^2}$, where
$c_2$ is the static charge, as follows:
\be
k_5 = {\overline c_2} \frac{\b}{N^2}\, ,\label{k5}
\ee
where ${\overline c_2}=c_2/a$, with $a$ the lattice spacing and $\b$ the lattice coupling.
The result, obtained from $SU(2)$ gauge theory on periodic and isotropic lattices, was compared to a holographic computation of the same quantity \cite{giatagan}.
There it was argued that this quantity, at least in the large $N$ limit, is in fact $N$-independent and this
could justify a comparison of the $N=2$ lattice result with the large $N$ gravity result.
The two calculations of $k_5$ showed a numerical agreement of $2\%$ \cite{5dCoulomb}, suggesting that
indeed, even if $k_5$ is $N$-dependent, its dependence on $N$ is probably rather weak.

In this work we extend the $SU(2)$ MF formalism of \cite{MF11,MF12} to $SU(N)$.
In \cite{Review} one can find expressions for the $SU(N)$ MF propagator on the isotropic lattice in the axial gauge.
Here we first extend the results of \cite{Review} by computing the propagator in a covariant gauge on the anisotropic lattice.
Moreover, we compute the Wilson Loop and the scalar observables, that have not been computed before for $SU(N)$.
As a first application, these quantities allow us to compute $k_5$ for large $N$ and verify its weak $N$-dependence
suggested by holography.

\section{Mean-field formalism}

The general formalism of the MF expansion was presented in detail in \cite{MF11,MF21} and we will not repeat it here.
For details one can consult these references together with \cite{Review}.
One is interested in physical observables, schematically written as
\be
\langle {\cal O} \rangle  =
\frac{1}{Z} \int \rD U \, {\cal O}[U] \re^{-S_W[U]} \, .
\ee
Here $S_W[U]$ is the usual Wilson plaquette action with anisotropic couplings $\b_4=\b /\g$
along four-dimensional hyperplanes and $\b_5=\b \g$ along the fiith dimension.
In this notation $\b$ is referred to as the lattice coupling and $\g$ as the anisotropy parameter.
The isotropic case is obtained for $\g=1$.
To first order in the MF expansion the expectation value of an observable takes the form
\be
\langle {\cal O} \rangle  =
{\cal O}[{\overline V}] + \frac{1}{2}{\rm tr}
\left\{\frac{\delta^2{\cal O}}{\delta V^2}\Biggr|_{\oV} K^{-1}\right\} \, ,\label{correction}
\ee
with
\be
K=-{K^{(hh)}}^{-1}+K^{(vv)}+K^{({\rm gf})}\label{Kdef}
\ee
the lattice propagator. $K^{(hh)}$ is the second derivative of the MF effective action with respect to
auxiliary degrees of freedom $H$ and $K^{(vv)}$ is the second derivative of the Wilson plaquette action
with respect to the link variables $V$. $K^{({\rm gf})}$ is the gauge fixing term.
${\cal O}$ is a gauge invariant operator and its second derivative with respect to the $H$-variables
contracted against the propagator and evaluated in the MF background defines an
expectation value.
The connected version of the latter is defined as ${\cal O}^c (t)= {\cal O} (t_0+t){\cal O} (t_0) $
and from it the correlator
\be
C (t) = <{\cal O}^c (t) > -
<{\cal O} (t_0+t)> <{\cal O} (t_0)>\,
\ee
is formed.
The ground state mass associated with the operator ${\cal O}$, to first order in the fluctuations, is then extracted from
\be
m = \lim_{t\to \infty} \ln \frac{C (t)}{C (t-1)}\, .
\ee

\subsection{Some $SU(N)$ integrals}

Here we review some facts related to $SU(N)$ integrals \cite{Review}.
Let $f(U)$ be a function of the $SU(N)$ link variables $U$ such that
$f(U)=f(VUV^{-1})$, $V\in SU(N)$.
Also let $\chi_r(U)$ be the character associated with the irreducible representation $r$ of $SU(N)$
that the group element $U$ is expressed in.
Then,
\be
f(U) = \sum_r f_r \chi_r(U), \hskip 1cm f_r = \int DU\, f(U) \chi^*_r(U)
\ee
A special case is the function
\be
e^{\frac{h_0}{N}{\rm Re}\, {\rm tr} (U)} = \sum_r c_r \chi_r(U)\, .
\ee
We express the irrep $r$ in terms of a set of positive integers as \cite{GS}
\be
r \longrightarrow \{\l^{(r )} \} = \{\l^{(r )}_1 \ge \l^{(r )}_2\ge \l^{(r )}_3 \ge \cdots \ge \l^{(r )}_N=0   \}\, .
\ee
We will need the $\l^{(r)}$ only for the symmetric and anti-symmetric representations. They are:
\bea
\l^{(2)} &=& (2,0,\cdots,0,{0})\nonumber\\
\l^{(1,1)} &=& (1,1,0,\cdots,0,{0})
\eea
respectively. Then,
\be
e^{\frac{h_0}{N}{\rm Re}\, {\rm tr} (U)} = \sum_{ \{\l^{(r )} \} }   \chi_{\{\l^{(r )} \}} (U) \sum_{n=-\infty}^{+\infty} {\rm det} \left\{
I_{{\l^{(r)}}_j-j+i+n}\left(\frac{h_0}{N}\right)\right\}\, ,
\ee
with $i,j=1,\cdots , N$ and $I_q$ the Bessel function of order $q$.
Using that
\be
\int DU  \chi_{\{\l^{(r)} \}} (U) = \d_{\{\l^{(r)}\},\{0\}}
\ee
we have
\bea
\int DU e^{\frac{h_0}{N}{\rm Re}\, {\rm tr} (U)} &=&
 \sum_{\{ \l^{(r)} \} } \d_{\{\l^{(r)} \},0} \sum_{n=-\infty}^{+\infty} {\rm det} \left\{
I_{{\l^{(r)}}_j-j+i+n}\left(\frac{h_0}{N}\right)\right\} \nonumber\\
&=& \sum_{n=-\infty}^{+\infty} {\rm det}\left\{ I_{-j+i+n}\left(\frac{h_0}{N}\right) \right\}\, .\label{z0}
\eea
An example is $N=2$ where
\be
\int DU e^{\frac{h_0}{2}{\rm Re}\, {\rm tr} (U)} = \sum_n {\rm det}\;
\begin{pmatrix}
I_n & I_{n-1}\cr
I_{n+1} & I_n
\end{pmatrix}\, .
\ee
In a matrix representation we can write a link in the irrep $\{\l^{(r)} \}$ as the matrix
\be
U(n',M') \longrightarrow D_{\a'\b'}^{\{\l^{(r)} \}}(n',M')\, .
\ee
The character of $\{\l^{(r)} \}$ can then be written as
\be
\chi_{\{\l^{(r)} \}}(U) = D_{\a\a}^{\{\l^{(r)} \}}\, .
\ee
We note the useful orthogonality relation: \be \int DU
D_{\a'\b'}^{\{\l^{(r)'}\}}D_{\a\a}^{\{\l^{(r)}\}} =
\d_{\{\l^{(r)}\},
\{\l^{(r)'}\}}\d_{\a'\b'}\frac{1}{d_{\{\l^{(r)}\}}}\, . \ee

\subsection{$K^{(hh)}$}

Let $U\in SU(N)$ the link (with explicit indices $U_{\a\b}(n,M)$)
and $H$ an $N\times N$ complex matrix. The indices $\a,\b =
1,\dots,N$ are the gauge indices, $n$ is the location of the link in
the lattice and $M$ is its direction. Define \bea
\zeta(H) &=& \int DU e^{\frac{1}{N}{\rm Re}\; {\rm tr}(UH) } \nonumber\\
u(H) &=& -\log {\zeta(H)}\; .
\eea
We are after derivatives of $u(H)$ with respect to $H_{\a'\b'}(n',M')$, evaluated in the
mean-field background
\be
V={\overline V} {\bf 1}, \hskip 1cm H={\overline H} {\bf 1}
\ee
with ${\bf 1}$ the $N\times N$ unit matrix.
We define the basic integral
\be
{\zeta_0} = \int DU e^{\frac{h_0}{N} {\rm Re} \; {\rm tr} (U)  }
=  \sum_{n=-\infty}^{+\infty} {\rm det}\left\{ I_{-j+i+n}\left(\frac{h_0}{N}\right) \right\}\, ,
\ee
where in a given parametrization ${\overline V}={\ov_0}$ and ${\overline H}={h_0}$.
We start from the first derivative of $u(H)$
\bea
&& u'_{\a'\b'}\equiv \frac{\partial
u(H)}{\partial H_{\b'\a'}(n',M')}\vert_{H={\overline H}{\bf 1}}=
-\frac{1}{\zeta} \frac{\partial \zeta(H)}{\partial H_{\b'\a'}(n',M')}\vert_{H={\overline H}{\bf 1}}\nonumber\\
&=& -\frac{1}{\zeta_0} \int DU \frac{\partial}{\partial H_{\b'\a'}(n',M')}
\left[\frac{1}{N} {\rm Re}\;{\rm tr} (UH) \right] e^{\frac{1}{N}{\rm Re}\; {\rm tr}(UH) }\vert_{H={\overline H}{\bf 1}}\nonumber\\
&=&  -\frac{1}{\zeta_0} \frac{1}{N} \ \zeta_1(\a',\b',n',M') =
-\frac{1}{\zeta_0} \frac{1}{N^2}{\zeta_1}\d_{\a'\b'}\,
,\label{uprime} \eea where \bea \zeta_1(\a',\b',n',M') &=&  \int DU
\, U_{\a'\b'}(n',M')\, e^{\frac{h_0}{N}{\rm Re}\; {\rm tr} (U) }
\equiv \d_{\a'\b'} \frac{\z_{1}}{N}\nonumber\\
\z_{1} &=& \int_{SU(N)} dU (Tr U) e^{\frac{h_0}{N}
Re Tr U}.\label{z1}
\eea
In the above expression we have dropped the argument $(n',M')$ from ${\zeta_1}_{\a'\b'}$ because the
background is uniform. In the anisotropic background the effect of the anisotropy is encoded in
the vev ${\overline H}$ (it will be $h_0$ along the $\m$-directions and $h_{05}$ along the fifth dimension).
The second derivative of $u(H)$ is
\bea
u_{\a',\b';\a'',\b''}'' &\equiv & \frac{\partial }{\partial H_{\b'\a'}(n',M')}\frac{\partial }{\partial H_{\b''\a''}(n'',M'')}u(H)
\vert_{H={\overline H}{\bf 1}}\nonumber\\
&=& \frac{1}{N^2 \zeta_0^2}\left[ \zeta_1(\a',\b',n',M')
\zeta_1(\a'',\b'',n'',M'') - \zeta_0
\zeta_2(\a',\b',n',M'; \a'',\b'', n'',M'') \right]\nonumber\\
\eea
with
\bea
&& \zeta_2(\a',\b',n',M'; \a'',\b'', n'',M'') = \nonumber\\
&& \int DU
\left[ \frac{\partial}{\partial H_{\b'\a'}(n',M')}
{\rm Re}\; {\rm tr} (UH) \right] \left[ \frac{\partial}{\partial H_{\b''\a''}(n'',M'')} {\rm Re}\; {\rm tr} (UH) \right]
e^{\frac{1}{N}{\rm Re}\; {\rm tr}(UH) }\vert_{H={\overline H}{\bf 1}}\nonumber\\
\eea
that is,
\be
 \zeta_2(\a',\b',n',M'; \a'',\b'', n'',M'') = \int DU\,  U_{\a'\b'} (n',M')\, U_{\a''\b''} (n'',M'')\,
 e^{\frac{h_0}{N}{\rm Re}\; {\rm tr} (U) }\, .\label{z2}
\ee
Next, we compute the integrals in
\eq{z1} and \eq{z2} respectively. We will make repeated use of the
notations: \be \left.  D_{n,N}(z) \equiv \det\left[
I_{n+j-i}\left(z\right)\right]_{1 \le i,j \le N},\ \ D_{n} \equiv
D_{n,N}(z)\right|_{z=\frac{h_0}{N}},\ee \be \left. \left. D_n^\prime
= \frac{d D_{n,N}(z)}{d z}\right|_{z=\frac{h_0}{N}} ,\ \ \
D_n^{\prime \prime}= \frac{d^2 D_{n,N}(z)}{d
z^2}\right|_{z=\frac{h_0}{N}}.\ee
A kind of generating function for $SU(N)$ has been
presented in \cite{carlsson}. It reads: \be
\left.G_{SU(N)}\right|_{\left(c,d\right)}\equiv \int_{SU(N)} dU e^{c
Tr U+d Tr U^\dagger} = \sum_{n=-\infty}^{+\infty}
\left(\frac{d}{c}\right)^\frac{n N}{2} \det\left[ I_{n+j-i}\left(2
\sqrt{c d}\right)\right]_{1 \le i,j \le N}.\ee
We note the following consequences:
\bea
\left. G_{SU(N)}\right|_{\left(\frac{h_0}{2 N},\frac{h_0}{2 N} \right)} &=&
\int_{SU(N)} dU e^{\frac{h_0}{N} Re Tr U} =
\sum_{n=-\infty}^{+\infty}
\det\left[I_{n+j-i}\left(\frac{h_0}{N}\right)\right]_{1 \le i,j \le
N} \equiv \zeta_0\nonumber\\
 \left. \frac{\partial G_{SU(N)}}{\partial c}
\right|_{\left(\frac{h_0}{2 N},\frac{h_0}{2 N} \right)} &=&
\int_{SU(N)} dU (Tr U) e^{\frac{h_0}{N} Re Tr U}  =
\sum_{n=-\infty}^{+\infty} \left[D_n^\prime -\frac{n N^2}{h_0} D_n
\right]\nonumber\\
\left. \frac{\partial G_{SU(N)}}{\partial d}
\right|_{\left(\frac{h_0}{2 N},\frac{h_0}{2 N} \right)} &=&
\int_{SU(N)} dU (Tr U^\dagger) e^{\frac{h_0}{N} Re Tr U}  =
\sum_{n=-\infty}^{+\infty} \left[D_n^\prime +\frac{n N^2}{h_0} D_n
\right]\nonumber\\
\left. \frac{\partial^2 G_{SU(N)}}{\partial c^2}
\right|_{\left(\frac{h_0}{2 N},\frac{h_0}{2 N} \right)} &=&
\int_{SU(N)} dU (Tr U)^2 e^{\frac{h_0}{N} Re Tr U} \nonumber\\
&=& \sum_{n=-\infty}^{+\infty} \left[D_n^{\prime\prime} -
\frac{N}{h_0}(1+2 n N) D_n^\prime + \frac{2 n
N^3}{h_0^2}\left(\frac{n N}{2}+1\right) D_n \right]\nonumber\\
\left. \frac{\partial^2 G_{SU(N)}}{\partial d^2}
\right|_{\left(\frac{h_0}{2 N},\frac{h_0}{2 N} \right)} &=&
\int_{SU(N)} dU (Tr U^\dagger)^2 e^{\frac{h_0}{N} Re Tr U} \nonumber\\
 &=& \sum_{n=-\infty}^{+\infty} \left[D_n^{\prime\prime} -
\frac{N}{h_0}(1-2 n N) D_n^\prime + \frac{2 n
N^3}{h_0^2}\left(\frac{n N}{2}-1\right) D_n \right]\nonumber\\
\left. \frac{\partial^2 G_{SU(N)}}{\partial c
\partial d} \right|_{\left(\frac{h_0}{2 N},\frac{h_0}{2 N} \right)}
&=& \int_{SU(N)} dU (Tr U) (Tr U^\dagger) e^{\frac{h_0}{N} Re Tr U}
= \sum_{n=-\infty}^{+\infty}
\left[D_n^{\prime\prime} + \frac{N}{h_0} D_n^\prime - \frac{n^2
N^4}{h_0^2} D_n \right]. \nonumber\\
\eea Since $D_n$ is even under $n\to -n$, all terms that are odd in
$n$ vanish in the sums. For the integral $\z_1({\a',\b',n',M'})$ we have 
\bea
\zeta_1(\a',\b',n',M') &=& \delta_{\a'\b'} \frac{1}{N} \int_{SU(N)} dU
[Tr U(n',M')] e^{\frac{h_0}{N} Re Tr U}  = \frac{\delta_{\a'\b'}
}{N} \z_{1},  \nonumber\\  
\z_{1}&=& \sum_{n=-\infty}^{+\infty}
\left[D_n^\prime -\frac{n N^2}{h_0} D_n \right]. 
\eea 
Thus, in this
notation, \be \z_0 = \sum_{n=-\infty}^{\infty} D_n, \hskip 1cm \z_1
= \sum_{n=-\infty}^{\infty} D_n'\; . \ee For the integral $\z_{2}$
we have to distinguish three different cases. The first is when we
take two derivatives with respect to $H'$ and $H''$, the second is
when we take one derivative with respect to $H'$ and one with
respect to $H''^{\dagger}$ and the third is when we take derivatives
with respect to $H'^{\dagger}$ and $H''^{\dagger}$. We call the
corresponding integrals as $\z_2^{00}$, $\z_2^{0+}$
($\z_2^{+0}=(\z_2^{0+})^\dagger$) and $\z_2^{++}(=\z_2^{00})$
respectively. We take all $H$'s in the fundamental representation.
Again, we can drop all space-time and directional arguments since
all the information about the background is contained in the
exponent of the integrand in \eq{z2}. We have \be
 \zeta_2(\a',\b',n',M'; \a'',\b'', n'',M'') = \int DU\,  U_{\a'\b'} (n',M')\,
 U_{\a''\b''} (n'',M'')\,
 e^{\frac{h_0}{N}{\rm Re}\; {\rm tr} (U) }
 \ee
 \be =c_1^{00} \delta_{\a'\b'} \delta_{\a''\b''}
 +c_2^{00} \delta_{\a'\b''} \delta_{\a''\b'}.
\ee We have set the integral equal to the two possible tensor
structures, with coefficients to be determined. Contracting with
$\d_{\a'\b'}\d_{\a''\b''}$ both sides we obtain \be \int DU\,
[\chi(U)]^2 e^{\frac{h_0}{N}{\rm Re}\; {\rm tr} (U) } =c_1^{00} N^2
+c_2^{00} N \equiv I_{\a\a;\b\b}, \ee while contracting with
$\d_{\a'\b''}\d_{\a''\b'}$ we obtain \be \int DU\, \chi(U^2)
e^{\frac{h_0}{N}{\rm Re}\; {\rm tr} (U) } =c_1^{00} N +c_2^{00} N^2
\equiv I_{\a\b;\b\a}. \ee Thus the coefficients can be expressed in
terms of the integrals: \be c_1^{00} =\frac{N I_{\a\a;\b\b} -
I_{\a\b;\b\a}}{N(N^2-1)},\ \ \ c_2^{00} =\frac{N I_{\a\b;\b\a}
-I_{\a\a;\b\b} }{N (N^2-1)}.\ee Following \cite{carlsson} we find
that \be I_{\a\a;\b\b} = \int DU\, [\chi_2(U) +\chi_{11}(U)]
e^{\frac{h_0}{N}{\rm Re}\; {\rm tr} (U) } \nonumber \ee \be  =
\sum_{n=-\infty}^{+\infty}
\det{I_{j-i+n+\l_i^{(2)}}\left(\frac{h_0}{N}\right)} +
\sum_{n=-\infty}^{+\infty}
\det{I_{j-i+n+\l_i^{(11)}}\left(\frac{h_0}{N}\right)},\ee \be
I_{\a\b;\b\a} = \int DU\, [\chi_2(U) -\chi_{11}(U)]
e^{\frac{h_0}{N}{\rm Re}\; {\rm tr} (U) } \nonumber \ee \be  =
\sum_{n=-\infty}^{+\infty}
\det{I_{j-i+n+\l_i^{(2)}}\left(\frac{h_0}{N}\right)} -
\sum_{n=-\infty}^{+\infty}
\det{I_{j-i+n+\l_i^{(11)}}\left(\frac{h_0}{N}\right)}\, . \ee These
integrals will be computed numerically. The $++$ case does not
require extra work, we have that $c_1^{++}=c_1^{00}$ and
$c_2^{++}=c_2^{00}$. The $0+$ case needs though separate
computation. The integral to be computed is \be
\z_2^{(0+)}(\a'\b';\a''\b'')\equiv J_{\a'\b'; \a''\b''} \equiv \int
DU\, U_{\a'\b'} (n',M')\, U_{\a''\b''}^\dagger (n'',M'')\,
e^{\frac{h_0}{N}{\rm Re}\; {\rm tr} (U) }\ee \be =c_1^{0+}
\delta_{\a'\b'} \delta_{\a''\b''} +c_2^{0+} \delta_{\a'\b''}
\delta_{\a''\b'}. \ee Following the same steps as before we find:\be
c_1^{0+} N^2+c_2^{0+} N = J_{\a \a; \b \b} = \int DU\, {\rm tr}(U)
{\rm tr}(U^\dagger) e^{\frac{h_0}{N}{\rm Re}\; {\rm tr} (U)
},\nonumber \ee \be c_1^{0+} N+ c_2^{0+} N^2 = J_{\a \b; \b \a} =
\int DU\, {\rm tr}(U U^\dagger) e^{\frac{h_0}{N}{\rm Re}\; {\rm tr}
(U) } = N \zeta_0, \ee which yields: \be c_1^{0+} = \frac{J_{\a \a;
\b \b}-\frac{1}{N} J_{\a \b; \b \a}}{N^2-1} = \frac{1}{N^2-1} \int D
U e^{\frac{h_0}{N}{\rm Re}\; {\rm tr} (U) } \left(Tr U Tr U^\dagger
-1\right),\ee \be c_2^{0+} = \frac{J_{\a \b; \b \a}-\frac{1}{N}
J_{\a \a; \b \b}}{N^2-1}= \frac{1}{N^2-1} \int d U
e^{\frac{h_0}{N}{\rm Re}\; {\rm tr} (U) } \left(N-\frac{1}{N} Tr U
Tr U^\dagger \right).\ee We can finally write down $K^{(hh)}$
directly in momentum space, since the Fourier transformation is
trivial (see \cite{MF11}). We introduce the index $q',q''=0,+$.
Then, \bea
&& K^{(hh)}_{M'M''}(p',\a',\b',q';p'',\a'',\b'',q'') \nonumber\\
&=& -\d_{p'p''} \frac{1}{N^2\z_0}\Bigl[ \z_2^{q'q''}(\a',\b';\a'',\b'')
-\frac{\z_1\d_{\a',\b'}\z_1\d_{\a'',\b''}}{N^2\z_0}\Bigr]_{{\overline v}_0,
{\overline h}_0}\cdot {\rm diag}(1,1,1,1,0)\nonumber\\
&-& \d_{p'p''} \frac{1}{N^2\z_0}\Bigl[ \z_2^{q'q''}(\a',\b';\a'',\b'')
-\frac{\z_1^2 \d_{\a',\b'}\d_{\a'',\b''}}{N^2\z_0}\Bigr]_{{\overline v}_{05},
{\overline h}_{05}}\cdot {\rm diag}(0,0,0,0,1)\nonumber\\
&=& -\d_{p'p''} \frac{1}{N^2\z_0}\Bigl[ (c^{q'q''}_1 \d_{\a'\b'}\d_{\a''\b''}
+ c^{q'q''}_2 \d_{\a'\b''}\d_{\a''\b'})-\z_{1}^2
\frac{\d_{\a',\b'}\d_{\a'',\b''}}{N^2\z_0}\Bigr]_{{\overline v}_0,
{\overline h}_0}\cdot {\rm diag}(1,1,1,1,0)\nonumber\\
&-& \d_{p'p''} \frac{1}{N^2\z_0}\Bigl[ (c^{q'q''}_1 \d_{\a'\b'}\d_{\a''\b''}
+ c^{q'q''}_2 \d_{\a'\b''}\d_{\a''\b'})-\z_{1}^2
\frac{\d_{\a',\b'}\d_{\a'',\b''}}{N^2\z_0}\Bigr]_{{\overline v}_{05},
{\overline h}_{05}}\cdot {\rm diag}(0,0,0,0,1)\, .\nonumber\\
\eea
The Euclidean indices $M'M''$ are shown explicitly in (diagonal) matrix form.
Before we continue, we will rewrite this expression in a more useful basis.
To begin, evaluating second derivatives in the mean-field background and leaving the
gauge index structure aside for the moment, corresponds to
\bea
&& (00): \;\;\;\; {\rm Re}\; [\partial_U \partial_U] = 1/4 (\partial^2_{u_0} - \partial^2_{u_A}) \nonumber\\
&& (0+): \;\;\;\; {\rm Re}\; [\partial_U \partial_{U^*}] = 1/4 (\partial^2_{u_0} + \partial^2_{u_A})
\eea
where the link is parametrized as $U=u_0 + i u_A$. Therefore the Hermitian (H) and anti-Hermitian (AH) channels
are
\bea
H &=& (00) + (0+)\nonumber\\
AH &=& (00) - (0+)\; .
\eea
On the other hand, in order to handle the gauge structure, one introduces the projectors
on the singlet and adjoint representations
\bea
P^{(S)} &=& \frac{1}{N} \d_{\a'\b'}\d_{\a''\b''}\nonumber\\
P^{(A)} &=& \d_{\a'\b''}\d_{\a''\b'}-\frac{1}{N} \d_{\a'\b'}\d_{\a''\b''}
\eea
and the new coefficients
\bea
c_a^H &=& c_{a}^{(00)} + c_{a}^{(0+)} \nonumber\\
c_a^{AH} &=& c_{a}^{(00)} - c_{a}^{(0+)}
\eea
with $a=1,2$ and
\bea
A^H &=&-\frac{1}{N^2\z_0} (c_2^H + N c_1^H - \frac{2\z_1^2}{N\z_0}), \hskip 1cm B^H = \frac{1}{N^2\z_0}  c_2^H\nonumber\\
A^{AH} &=&-\frac{1}{N^2\z_0} (c_2^{AH} + N c_1^{AH}), \hskip 1cm B^{AH} = \frac{1}{N^2\z_0}  c_2^{AH}\label{coeff}
\eea
Now we can write the non-vanishing components of $K^{(hh)}$ simply as
\bea
K^{(hh)}_{\m'\m''}(p',\a',\b';p'',\a'',\b''; z) &=& \d_{p'p''} \Bigl(A^z P^{(S)}_{\a'\b';\a''\b''} \oplus B^z P^{(A)}_{\a'\b';\a''\b''}\Bigr)_{{\overline v}_0,
{\overline h}_0}\; \d_{\m'\m''}\; \nonumber\\
K^{(hh)}_{55}(p',\a',\b';p'',\a'',\b''; z) &=& \d_{p'p''} \Bigl(A^z P^{(S)}_{\a'\b';\a''\b''} \oplus B^z P^{(A)}_{\a'\b';\a''\b''}\Bigr)_{{\overline v}_{05},
{\overline h}_{05}}\;
\eea
where $z=H,AH$.
One can check that for any $N$
\be
B^{AH} = \frac{2u'}{N{\overline h}_0} =  -\frac{2\ov_0}{N{\overline h}_0}=-\frac{2}{xN^3}\frac{\z_1(x)}{\z_0(x)}\, .\label{cond}
\ee
We have defined $x=\overline h_0/N$ and used the background solution given later in \eq{bsol}.

For large $N$, the numerical computation of the determinants
involved in the coefficients is plagued by instabilities due to
large cancellations. It is then useful to consider the large $H$
expansion of $u(H)$ \cite{Review} \bea u(H) &=& \ln
\left[{\frac{\prod_1^{N-1}k!}{(2\pi)^{(N-1)/2}}}\right] +
\frac{N^2-2}{2}\ln{N} + {\hat H} -\frac{N^2-1}{2}\ln{\hat H} -
\frac{N^2-1}{8{\hat H}}+\cdots \eea with ${\hat H}=\frac{{\rm tr}
\{H\}}{N}$, which is equal to ${\overline h}_0$ in the mean-field
background. Noticing in addition that as $N$ increases $\b_c$ and
${\overline h}_0$ also increase we can conclude that already for
relatively low values of $N$ ($N=5,6,\cdots$) the asymptotic
expansion of $u(H)$ is a good approximation.

In this limit we can use the approximate expressions
\bea
A^H &\simeq& -\frac{N}{{\overline h}_0^2}, \hskip 2.7cm B^H \simeq 0 \nonumber\\
A^{AH} &\simeq& 0, \hskip 2.7cm B^{AH} = -\frac{2{\ov }_0}{N{\overline h}_0}\label{asymptcoeff}
\eea
and for later reference
\be
{\ov }_0 = \frac{\z_1}{N\z_0}\simeq 1-\frac{N^2-1}{2 {\overline h}_0}\, .\label{asymptMF}
\ee
For $N=2$, $B^H$ and $A^{AH}$ are identically zero and for large $N$, they become negligible \cite{Review}.
Notice also that the expression for $B^{AH}$ is exact for any $N$, see \eq {cond}.
For instance, on the isotropic torus we have
\be
B^{AH} = -\frac{1}{4} \frac{1}{N\beta {\ov}_0^2}\, .
\ee
When ${\overline h}_0\to \infty$ we have $\ov_0\simeq 1$ and we obtain the asymptotic expression
$B^{AH} = -\frac{2}{N{\overline h}_0}$.

\subsection{$K^{(vv)}$}

The space-time structure of $K^{(vv)}$ for $SU(N)$ is similar to the $SU(2)$ case.
In $q'-q''$ space it is a two by two matrix, just like $K^{(hh)}$. The hermitian (anti-hermitian) channel of the
$SU(N)$ model corresponds to the hermitian (anti-hermitian) channel of the $SU(2)$ model.
The gauge index structures of $K^{(vv)}$ and of the gauge fixing term can be seen from
the second derivative of the Wilson plaquette action (suppressing all but the group indices)
\be
\frac{\partial^2 {\rm tr} \left[U_{\a\b}U_{\b\g}U^\dagger_{\g\d}U^\dagger_{\d\a}\right]}{\partial U_{\a'\b'}\partial U_{\a''\b''}}
\sim \d_{\a'\b''}\d_{\a''\b'}N
\ee
and the gauge fixing term
\be
\frac{\partial^2 {\rm tr} \left[U_{\a\b}U_{\b\a}\right]}{\partial U_{\a'\b'}\partial U_{\a''\b''}}
\sim \d_{\a'\b''}\d_{\a''\b'}N
\ee
respectively.

We define the bond shifting operators
\bea
&&{\Delta}_{AH}=-(N\frac{\b}{\g}{\ov}_0^2)\cdot \nonumber\\
&& \left(\begin{array} {cc} \sum^\prime c_{M'}-\frac{1}{\xi} s^2_{0/2} \hskip 1cm  ys_{0/2}s_{1/2}
 \hskip 1cm ys_{0/2}s_{2/2} \hskip 1cm ys_{0/2}s_{3/2}  \hskip 1cm y_5s_{0/2}s_{5/2}
\g^2 \frac{\ov_{05}}{{\ov}_0} \\
ys_{1/2}s_{0/2}  \hskip 1cm \sum^\prime c_{M'}-\frac{1}{\xi}  s^2_{1/2}\hskip 1cm
ys_{1/2}s_{2/2} \hskip 1cm ys_{1/2}s_{3/2} \hskip 1cm y_5s_{1/2}s_{5/2}\g^2
\frac{\ov_{05}}{{\ov}_0} \\
ys_{2/2}s_{0/2} \hskip 1cm ys_{2/2}s_{1/2} \hskip 1cm
\sum^\prime c_{M'}-\frac{1}{\xi}  s^2_{2/2}\hskip 1cm
ys_{2/2}s_{3/2}   \hskip 1cm y_5s_{2/2}s_{5/2}\g^2 \frac{\ov_{05}}{{\ov}_0}\\
ys_{3/2}s_{0/2}  \hskip 1cm ys_{3/2}s_{1/2} \hskip 1cm
ys_{3/2}s_{2/2} \hskip 1cm \sum^\prime c_{M'}- \frac{1}{\xi}  s^2_{3/2}\hskip 1cm
y_5s_{3/2}s_{5/2}\g^2 \frac{\ov_{05}}{{\ov}_0} \\
y_5s_{5/2}s_{0/2} \g^2 \frac{\ov_{05}}{{\ov}_0} \hskip .5cm  y_5s_{5/2}s_{1/2} \g^2 \frac{\ov_{05}}{{\ov}_0}\hskip .5cm
y_5s_{5/2}s_{2/2} \g^2 \frac{\ov_{05}}{{\ov}_0}\hskip .5cm  y_5s_{5/2}s_{3/2}
\g^2 \frac{\ov_{05}}{{\ov}_0} \hskip .5cm
\g^2 \left(\sum^\prime c_{M'}- \frac{1}{\xi}  s^2_{5/2}\right)\\
\end{array} \right)\nonumber\\ \label{Kvva1}
\eea
where
\be
y=2-1/\xi. \hskip 1cm y_5=2\g^2\frac{\ov_{05}}{\ov_0}-\frac{\g}{\xi}
\ee
and
\bea
&&{\Delta}_H =-(N\frac{\b}{\g}{\ov}_0^2)\cdot\nonumber\\
&&\left(\begin{array} {cc} \sum^\prime c_{M'} \hskip 1cm  2c_{0/2}c_{1/2}
 \hskip 1cm 2c_{0/2}c_{2/2} \hskip 1cm 2c_{0/2}c_{3/2}  \hskip 1cm 2c_{0/2}c_{5/2}
\g^2 \frac{\ov_{05}}{{\ov}_0} \\
2c_{1/2}c_{0/2}  \hskip 1cm \sum^\prime c_{M'} \hskip 1cm
2c_{1/2}c_{2/2} \hskip 1cm 2c_{1/2}c_{3/2} \hskip 1cm 2c_{1/2}c_{5/2}\g^2
\frac{\ov_{05}}{{\ov}_0} \\
2c_{2/2}c_{0/2} \hskip 1cm 2c_{2/2}c_{1/2} \hskip 1cm
\sum^\prime c_{M'} \hskip 1cm
2c_{2/2}c_{3/2}   \hskip 1cm 2c_{2/2}c_{5/2}\g^2 \frac{\ov_{05}}{{\ov}_0}\\
2c_{3/2}c_{0/2}  \hskip 1cm 2c_{3/2}c_{1/2} \hskip 1cm
2c_{3/2}c_{2/2} \hskip 1cm \sum^\prime c_{M'} \hskip 1cm
2c_{3/2}c_{5/2}\g^2 \frac{\ov_{05}}{{\ov}_0} \\
2c_{5/2}c_{0/2} \g^2 \frac{\ov_{05}}{{\ov}_0} \hskip .5cm  2c_{5/2}c_{1/2} \g^2 \frac{\ov_{05}}{{\ov}_0}\hskip .5cm
2c_{5/2}c_{2/2} \g^2 \frac{\ov_{05}}{{\ov}_0}\hskip .5cm  2c_{5/2}c_{3/2}
\g^2 \frac{\ov_{05}}{{\ov}_0} \hskip .5cm
\g^2 \sum^\prime c_{M'}\\
\end{array} \right)\, .\nonumber\\
\eea We use the notation $s_{0/2}=\sin{p'_0/2}$ etc. and $c_5 = \g^2
\frac{{\ov}_{05}^2}{{\ov}_{0}^2} \cos{(p_5')} $. These expressions
appear in \cite{Review} in the axial gauge and for isotropic
lattices. Here we have generalized them to anisotropic lattices and
we have fixed a covariant gauge parametrized by $\xi$ as in
\cite{MF11}. We have checked numerically that our results do
not depend on the value of $\xi.$ Using $\d_{\a'\b''}\d_{\a''\b'} =
P^{(S)} + P^{(A)}$ we can express ${K}^{(vv)}$ as \bea
{K}^{(vv)}_{M'M''}(p',\a',\b';p'',\a'',\b'';H) &=& \d_{p'p''} (P^{(S)}_{\a'\b';\a''\b''} \oplus P^{(A)}_{\a'\b';\a''\b''}) {{\Delta}_{H}}_{M'M''} \nonumber\\
{K}^{(vv)}_{M'M''}(p',\a',\b';p'',\a'',\b'';AH) &=&  \d_{p'p''} (P^{(S)}_{\a'\b';\a''\b''} \oplus P^{(A)}_{\a'\b';\a''\b''}) {{\Delta}_{AH}}_{M'M''} \, .
\eea
The $SU(N)$ propagator is then
\be
K_{M'M''} = - {K^{(hh)-1}_{M'M''}} + K^{(vv)}_{M'M''}\, .
\ee
The notation here and from now on is that $K^{-1}_{M'M''}$ represents the $M'M''$'th element of $K^{-1}$.
The inverse of $K$ has four sectors, labeled by $z=H,AH$ and $w=S,A$.
Schematically we represent these sectors as $K^{-1}(z,w)$
and express them in this basis as
\be
K^{-1}(z,w) = K^{-1}(H,S) P^{(S)} +  K^{-1}(H,A) P^{(A)}+K^{-1}(AH,S) P^{(S)} +  K^{-1}(AH,A) P^{(A)}\, .
\ee

\subsection{The phase diagram}

In the anisotropic theory, there are two mean values for the links,
$\ov_0$ and $\ov_{05}$ determined by the extremization of ($d=5$)
\be \frac{S_{\rm eff}[{\overline V},{\overline H}]}{{\cal N}}= -\b_4
\frac{(d-1)(d-2)}{2} \ov_0^4 - \b_5 (d-1) \ov_0^2 \ov_{05}^2 + (d-1)
u({\overline h}_0) + u({\overline h}_{05}) + (d-1) {\overline h}_0
\ov_0 +{\overline h}_{05} \ov_{05}\, .\label{MF} \ee Here ${\cal N}
= L^3 N_5$ is the number of spatial lattice points with $N_5$ the
number of points in the fifth dimension. \eq{MF} yields conditions
identical in form to the $SU(2)$ case \cite{MF11} \bea && \ov_0 = -
u({\overline h}_0)'\,, \hskip 1cm
{\overline h}_0=\frac{6\b}{\g}\ov_0^3+2\b\g \ov_0 \ov_{05}^2 \,,\nonumber\\
&& \ov_{05} = - u({\overline h}_{05})'\,, \hskip 1cm
{\overline h}_{05}=8\b\g\ov_0^2 \ov_{05} \,,\label{MF_a}
\eea
where $u'$, according to \eq{uprime} is (here we differentiate form $N=2$)
\be
u' = {\rm tr} \{u_{\a'\b'} \} =  -\frac{1}{\z_0N^2}N \z_{1}\, ,
\ee
that is
\be
\ov_0 = \frac{\z_1(x)}{N\z_0(x)}\; .\label{bsol}
\ee
For $N=2$ this reduces to the known result
$\ov_0 = I_2(2x)/I_1(2x)$.

One can immediately see that when all momenta vanish, in the $(AH,A)$ sector we have
(for simplicity we take $\g=1$), using \eq{cond}:
\be
K(AH,A) = -\left(\frac{N{\overline h}_0}{2u'} + 4N\b \ov_0^2\right)\cdot {\bf 1}
\ee
which vanishes by \eq{MF_a}.  These are the five expected torons, which persist even when $\g\ne 1$.
The various phases on the phase diagram are defined as follows:
\begin{itemize}
\item Confined phase: $\ov_0 =0, \ov_{05}=0$
\item Coulomb phase: $\ov_0 \ne 0, \ov_{05}\ne 0$
\item Layered phase: $\ov_0 \ne 0, \ov_{05}=0$
\end{itemize}
We will perform our computations in the Coulomb phase where the background is non-vanishing everywhere.
In addition we will stay near the phase transition where we expect that cut-off effects are suppressed.
The critical value of the coupling $\b_c$ that signals the end of the Coulomb phase
moves towards to larger values as $N$ increases. Apart from that, the phase diagram is
qualitatively similar to the $SU(2)$ phase diagram as described in \cite{MF11}.

\section{Observables}

To first order in the fluctuations we will be computing
\be
\langle {\cal O} \rangle = {\cal O}[\overline V] +\frac{1}{2} {\rm tr}
\Bigr\{ \sum_{w=S,A}\sum_{z=H,AH}\Bigl( \frac{\d^2 {\cal O}} {\d V^2} \Bigr)_{V=\ov_0}(z,w) (K^{-1})(z,w)\Bigr\}\, .\label{1st}
\ee
We will often call the above second derivative of the observable simply as ${\cal O}_2$ which also
can be expressed as
\be
{\cal O}_2(z) = {\cal O}_2(z,S) P^{(S)} + {\cal O}_2(z,A) P^{(A)} \, .
\ee
Then we can expand as
\bea
\langle {\cal O} \rangle &=& {\cal O}[\overline V] +\frac{1}{2}\sum_{z,w} {\rm tr} \{{\cal O}_2(z,w) K^{-1}(z,w)\}\nonumber\\
&=& {\cal O}[\overline V] + \nonumber\\
&& \frac{1}{2}{\rm tr} \left[{\cal O}_2(H,S) K^{-1}(H,S) + {\cal O}_2(AH,S) K^{-1}(AH,S)\right] {\rm tr} \{\bf 1\}_S\nonumber\\
&& \frac{1}{2}{\rm tr} \left[{\cal O}_2(H,A) K^{-1}(H,A) + {\cal O}_2(AH,A) K^{-1}(AH,A)\right] {\rm tr} \{\bf 1\}_A\nonumber\\
&=& {\cal O}[\overline V] + \nonumber\\
&& \frac{1}{2}{\rm tr} \left[{\cal O}_2(H,S) K^{-1}(H,S) + {\cal O}_2(AH,S) K^{-1}(AH,S)\right] N \nonumber\\
&& \frac{1}{2}{\rm tr} \left[{\cal O}_2(H,A) K^{-1}(H,A) + {\cal O}_2(AH,A) K^{-1}(AH,A)\right] N (N^2-1)\, .\label{sectors}
\eea

\subsection{The scalar mass}

The Euclidean structure is identical to that of the $SU(2)$ model.
We recall the $SU(2)$ result (for $SU(2)$ the gauge index takes the values
$\a'=0,1,2,3$ where $\a' =0$ represents the Hermitian-Singlet channel and $\a' =1,2,3$ the anti-Hermitian-Adjoint channel)
where only the Hemitian channel, free of zero modes, is present \cite{MF11}
\be
C_S(t) = \frac{2}{\cal N} (P_0)^2 \sum_{p_0'}\cos(p_0't)\sum_{p_5'}
|\D^{(N_5)}(p_5')|^2 K_{55}^{-1}\left((p_0',{\vec 0},p_5'),\a'=0;(p_0',{\vec 0},p_5'),\a'=0\right)\; .
\ee
Recall that for $SU(2)$ for any observable there are at most the two sectors, $(H,S)$ and $(AH,A)$ contributing.

For $SU(N)$ the contribution of the adjoint channel vanishes identically for basically the same reason as for $SU(2)$, that is
because taking only one derivative of the observable with respect to a fluctuation along the Lie algebra and evaluating it in the background
gives ${\cal O}_2 (H,A)={\cal O}_2 (AH,A)=0$ in \eq{sectors}.
Also, since the double derivative on the Polyakov loop that represents the scalar has the gauge index structure
\be
\d_{\a'\b''}\d_{\a'',\b'} = P^{(S)} + P^{(A)}
\ee
and the schematic structure
\bea
C_S(t)  \sim N {\cal O}_2 (H,S) K^{-1}(H,S) +  N {\cal O}_2 (AH,S) K^{-1}(AH,S) \, ,
\eea
the final expression for the correlator becomes
\bea
C_S(t) =\frac{NN_5^2}{\cal N} \frac{(P_0)^2}{v_{05}^2} \sum_{p_0'}\cos(p_0't)\d_{p'_k,0}\d_{p'_5,0}\Bigl[K_{55}^{-1}\left(p';p';H,S\right) +
K_{55}^{-1}\left(p';p';AH,S\right)\Bigr] \nonumber\\
\eea
where $k=1,2,3$.
Notice that for $SU(2)$ the coefficient $A^{AH}$ vanishes identically, so only the $(H,S)$ sector contributes,
consistently with our comment above.
We have defined
\be
\D^{(N_5)}(p) =  \frac{1}{v_{05}} \sum_{r=0}^{N_5-1} e^{ip(r+1/2)} = N_5 \frac{\d_{p_5,0}}{v_{05}}\, ,
\ee
where $r=0$ labels the first link along the fifth dimension, $r=1$ the second, etc.
$P_0$ is the mean-field length of the Polyakov loop representing the scalar observable.
It is easy to see that the toron $p_0=0$ adds a zero to the scalar correlator since it
contributes a time independent constant. Given that the Hermitian channel of the propagator does not
contain any zero modes, there is no toron.
The plateau in the time decay of this observable yields the mass of the mass of the scalar observable $a_4m_S$ in lattice units.

The vector mass is a purely finite volume effect (in particular it does not depend on $N$) so we can use directly the $SU(2)$ relation
\cite{MF11,MF12}
\be
a_4 m_V = \frac{4\pi}{L}\; .
\ee

\subsection{The Wilson Loop}

Here we compute the Wilson Loop (WL) along four-dimensional hyperplanes.
Again, the space-time structure is the same as for $SU(2)$, see \cite{MF11}.

For an observable represented by a single loop it makes a difference
whether the two links that are removed (by the two derivatives) are pointing in the same or
opposite directions. In the former case there is a relative minus sign with respect to the latter.
A case where this remark applies is the Wilson Loop.  The schematic structure in this case is
\bea
{\rm exchange:} && (00) +  \, (0+)\nonumber\\
{\rm tadpole:} && (00) -  \, (0+)
\eea
corresponding to the two diagrams that contribute to leading non-trivial order, an exchange
between the two spatial legs and the tadpole on a given spatial leg.
For general $N$, in principle, all four sectors contribute.
The final result is:
\bea
V_5(r) &=& -\frac{1}{\ov_0^2}\frac{N}{2\cal N}
\sum_{k=1}^3 \sum_{p'} \d_{p_0',0}\cdot \nonumber\\
&&\Biggr\{\Bigl[ (\frac{1}{3}\cos(p_k'r)+1)K_{00}^{-1}(p';p';H,S)+ (N^2-1)(\frac{1}{3}\cos(p_k'r)-1) K_{00}^{-1}(p';p';AH,A) \Bigr]\nonumber\\
&& \Bigl[(\frac{1}{3}\cos(p_k'r)-1)K_{00}^{-1}(p';p';AH,S)+ (N^2-1)(\frac{1}{3}\cos(p_k'r)+1) K_{00}^{-1}(p';p';H,A)\Bigr]\Biggr\}\nonumber\\
\label{StaticTorusa4d} \eea where we have dropped the irrelevant
additive constant originating from the zeroth order contribution.

For $SU(2)$ only the first two sectors contribute.
In particular, the two terms in the last line vanish because $A^{AH}=B^H=0$.
Finally the toron in the propagator in the $(AH,A)$ sector is cancelled by the zero in the observable
in the same sector since the factor $\cos(p_k'r)-1$ vanishes for $p=0$.
Notice that for $SU(2)$ the coefficient $A^{AH}$ is not being used by neither the scalar nor by the WL.

\section{A simple application: Coulomb's constant in five dimensions}
\label{holo}

We present a simple application of the formalism we developed, on
the isotropic lattice. As mentioned in the Introduction, the
dimensionless quantity $k_5$ defined in \eq{k5} is computable both
in lattice gauge theory and in gravity. An agreement between the two
computations would constitute evidence for the validity of the
holographic conjecture \cite{Mald}. We remind  that the holographic
computation of $k_5$ defined in \eq{k5} yields
\cite{giatagan,5dCoulomb} (expressed in lattice parameters) the
number \be k_5  =
\left[\frac{B(2/3,1/2)}{3\pi^{2/3}}\right]^3=0.0649    \label{k5G}
\ee with $B(x,y)$ the Euler Beta function, for any $N$ for which the
computation is valid. This is expected to be the case in the large
$N$ limit, therefore any agreement with the lattice computation of
the same quantity is expected to occur at least in this limit.

From the point of view of the lattice, we define trajectories on the phase diagram that approach the bulk
first order phase transition, with
\be
q = \frac{a_4 m_V}{a_4m_S}\label{q}
\ee
kept constant \cite{MF12}.
The quantity $k_5={\overline c}_2{\beta}/{N^2}$ on the isotropic lattice depends on $\b$ and $L$.
Using these trajectories, it is however straightforward to obtain its infinite $L$ value via extrapolation.
Here we choose $q=2$ even though the
infinite $L$ extrapolations of $k_5$ are expected to be essentially $q$-independent.

\begin{figure}[!t]
\centerline{\epsfig{file=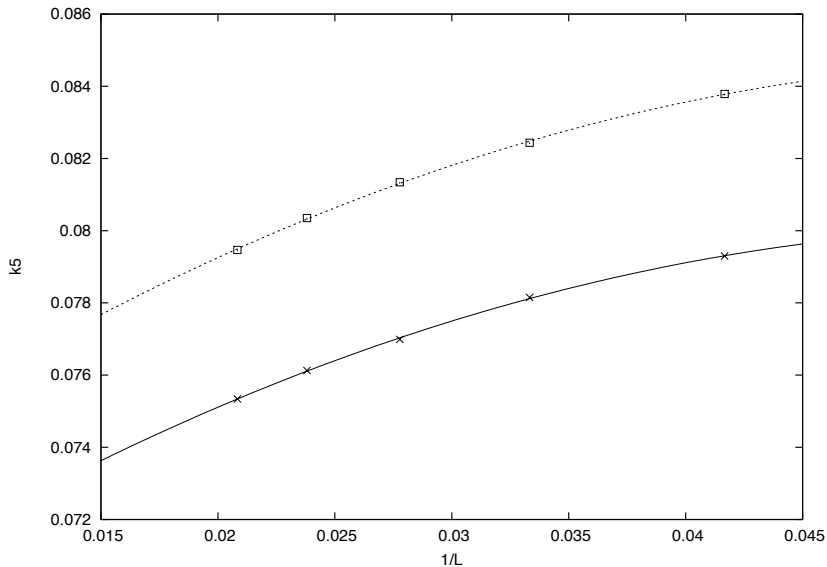,width=12cm}}
\caption{\small
Comparison of asymptotic expressions (lower curve) versus
sum over determinants (upper curve) for SU(5).
\label{versus}}
\end{figure}

In figure \ref{versus} we compare for $SU(5)$ the results obtained
from the asymptotic expansions \eq{asymptcoeff} for the coefficients
in $K^{(hh)}$ together with the asymptotic form for the background
\eq{asymptMF} against the ones obtained by the sum over determinants
\eq{coeff}. We expect that the former (latter) would be more
reliable for large (small) values of $N$ and indeed for $N=5$ the
discrepancy between the two is already quite small. According to
this result, for $N\le 5$ we will be computing the coefficients
using the sum over determinants method and for $N>5$ using their
asymptotic expressions.

\begin{figure}[!t]
\centerline{\epsfig{file=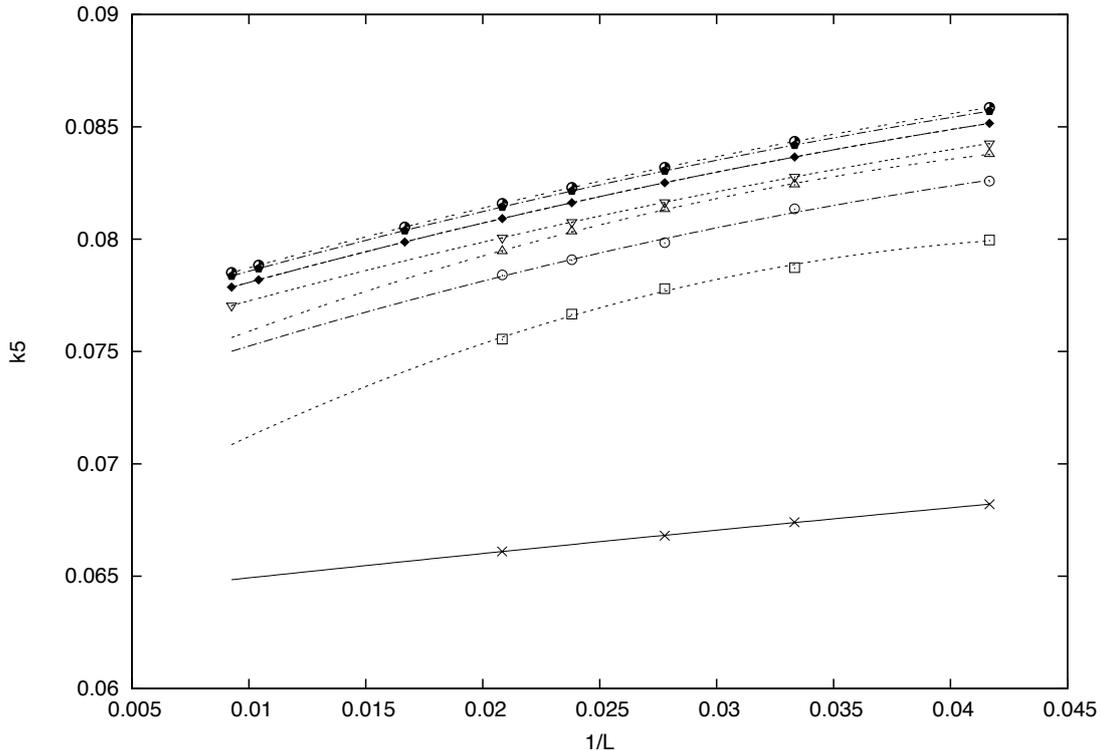,width=16cm}}
\caption{\small
$k_5$ versus ${1}/{L}$ for $SU(2)$ (lowest curve), $SU(3)$, $SU(4)$, $SU(5)$,
$SU(10)$, $SU(15)$, $SU(30)$ and $SU(100)$.
\label{all}}
\end{figure}

In figure \ref{all} we show the values obtained for $k_5$ for increasingly large values of $N$.
For $N=2,3,4,5$ we use the sum over determinants expressions
and for $N>5$ we use their asymptotic forms.
We also show quadratic fits to
the data. We observe that the intercept increases slightly with $N.$
The largest value depicted is $N=100$.
Data for larger $N$ than that become indistinguishable from the latter with naked eye.
It is clear that the curves saturate for large $N$.
This demonstrates that $k_5$ becomes $N$-independent as $N\to \infty$.

\begin{figure}[!t]
\centerline{\epsfig{file=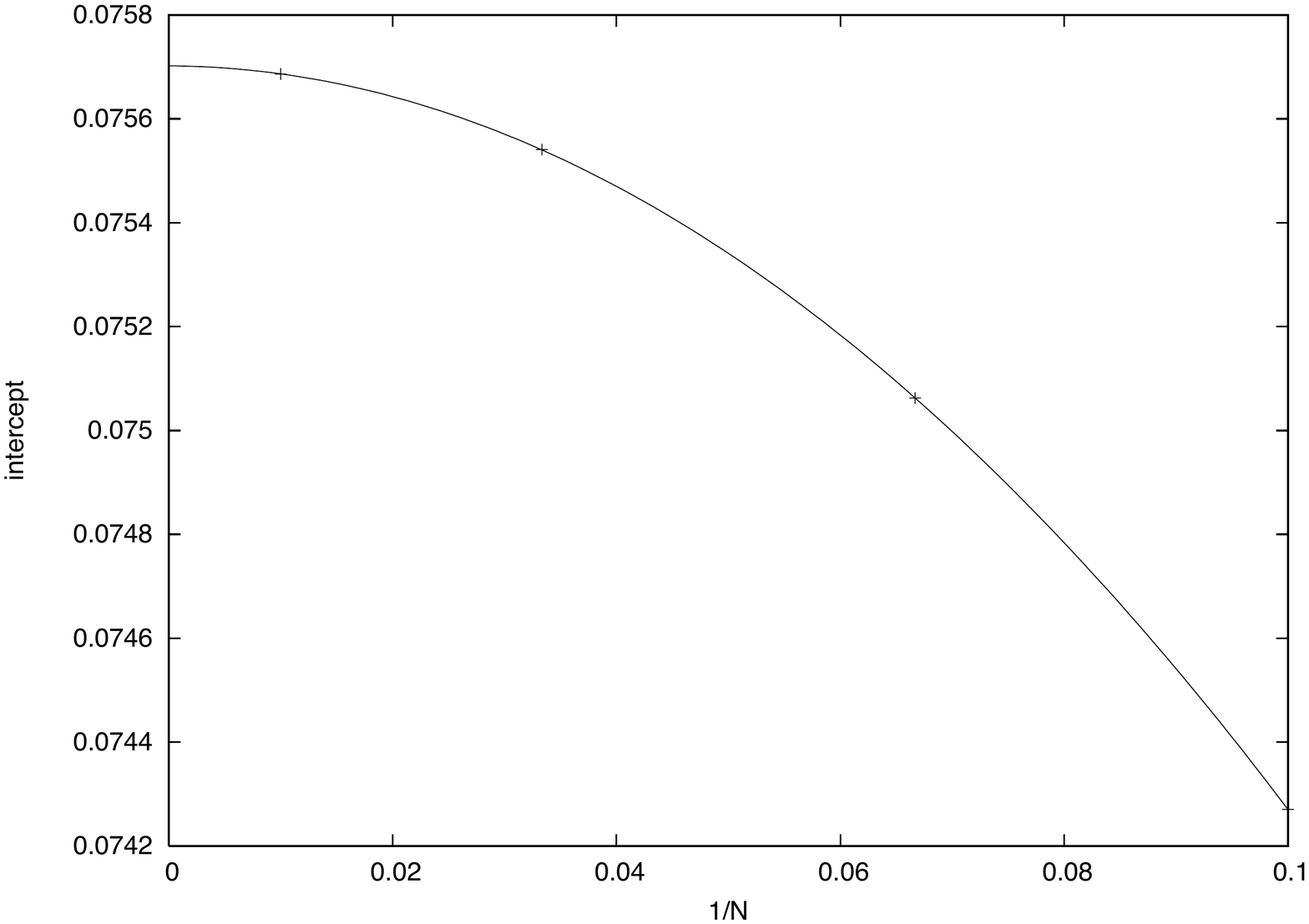,width=16cm}}
\caption{\small
Intercept at $L=\infty$ versus ${1}/{N}$ along with a
quadratic fit, predicting the value $k_5=0.0757$ for $SU(\infty).$
\label{intercept}}
\end{figure}

In figure \ref{intercept} we depict the $N$ dependence of the
intercept for $N\ge 10$ along with a quadratic fit. It cuts the
vertical axis at $k_5=0.0757,$ presumably the result for $SU(\infty).$
This should be compared against the holographic result $0.0649$,
resulting into a $17\%$ disagreement.
Evidently the $N=\infty$ value of $k_5$ is farther away compared to its $N=2$ value
which is just $2\%$ away from the holographic result.
This "reverse" trend could be related to the order of the computations
on both the gauge theory and gravity sides.
The essential fact to keep in mind is that the mean-field prediction for $k_5$ in the large $N$ limit stabilizes
to a value not too far away from the holographic prediction.

%
%

%

\section{Discussion}

In order that the comparison between the Mean-Field and Holographic computations makes sense,
both should be reasonably good descriptions of the 5d $SU(N)$ gauge theory at intermediate to strong coupling.
Assuming that from the holographic point of view this is a conjecture and provided that the Mean-Field
is a good analytical tool at strong coupling, any agreement
with the gravity side gives evidence for the validity of the holographic conjecture.
Reversely, if we assume that the holographic duality holds, then any observable
derived from the classical gravitational system should be in principle verifiable from the
gauge theory side, provided that it can be computed reliably at the non-perturbative level.
No matter what point of view is taken, a necessary condition is that the 
truncated Mean-Field expansion we have employed here is a good approximation to the non-perturbative system.
We have mentioned in the Introduction some general arguments to this effect, such as the
validity of the expansion improving as the dimensionality of the system increases.
The only thing that can decide about this issue is a comparison with
Monte Carlo simulations. In fact, it is known for some time that the Mean-Field predicts the
order of the bulk phase transition and even the numerical value of the associated critical
coupling correctly \cite{Review}. Note that this is a prediction already at the level of the MF background, i.e. without
taking into account fluctuations. 
In \cite{FrancMagdAnt} a more sophisticated comparison between the MF and Monte Carlo methods
was performed on the anisotropic $N=2$ lattice.
The MF predicts, via the Wilson Loop (which can be obtained in the MF only
by taking into account fluctuations), dimensional reduction via localization for $\g\simeq 0.55$ \cite{MF11,MF12}.
This was verified by MC simulations that showed \cite{FrancMagdAnt} that in the same regime of the anisotropy, 
Polyakov Loops fluctuate independently along four-dimensional hyperplanes.
In \cite{MF21} the mass of observables with 4d scalar and vector quantum numbers are computed
with orbifold boundary conditions along the fifth dimension
(these however should not interfere with the issue of the convergence of the MF expansion).
Also these observables can be computed with the MF method only in the presence of fluctuations.
The agreement was seen to be again good. All these results give us 
an increasing confidence that the MF is indeed a good description of the 5d system at the non-perturbative level.

More specifically now, the computation of $k_5$ from the gravity side suggested that the dual gauge theory,
at leading order in a large $N$ expansion 
a) must be not weakly coupled by the nature of the duality
b) is in its 5d Coulomb phase since the static potential is of a $1/r^2$ form
c) sits in a regime of its phase diagram where cut-off effects are suppressed since in the result there is no
sign of the presence of a gauge theory regulator.
The first observation forced us to move away from the 5d perturbative point where the coupling
is expected to go to zero when the cut-off is removed. Moving into the interior of the phase diagram
leads us eventually on the bulk phase transition because everywhere else physics is heavily
cut-off dominated. The second observation on the other hand prohibits us from crossing the phase transition.
The only choice we are left with is to be as near as possible to the phase transition, which is
precisely where we have computed $k_5$ from the gauge theory side.
What is left to be addressed is why do we expect cut-off effects to be suppressed near the phase transition,
or in lattice language, why do we generally expect the lattice spacing to decrease as $\b \to \b_c$. 
A milder version of this question, related to our quantity $k_5$ is what is its
lattice spacing dependence near the phase transition.
It is not easy to answer definitively the former version of the question.
What one can observe though from the MF computation \cite{MF11} is that the scalar mass
in units of lattice spacing $a m_S$ decreases as the phase transition is approached.
It is however a first order phase transition therefore most likely the lattice spacing remains finite and in fact
one can see that $am_S$ can be pushed down to approximately 0.1 but not much further.
This means that finite lattice spacing effects will be present even at the phase transition, even though not very large.
Regarding $k_5$, it has been defined \cite{5dCoulomb} by analogy to the non-abelian Coulomb's constant in four dimensions,
where one can see that by construction it is a cut-off independent quantity. Here, our working assumption was  
that the lattice spacing dependence in $k_5$ (via the product ${\overline c_2}\b$) cancels at least near the phase transition,
just like in 4d and according to what the gravitational result suggests.
The fact that as we move closer to the phase transition the value of $k_5$ changes by very little, especially for large $N$,
strongly supports this assumption. Notice that when $L$ is increased while keeping $q$ in \eq{q} constant, takes us
closer to the phase transition and thus changes the lattice spacing.

We now discuss other possible sources for the observed 17$\%$ discrepancy.
From the gravity side, in infinite four-dimensional volume, there may be $\a'$
and finite $N$ corrections to the static potential. It would be interesting to see if it is possible to compute 
such corrections for the $D4$-brane background, which is the basis for the holographic computation of sect. \ref{holo}.
From the gauge theory side we also expect to have corrections.
Already at leading order in the MF expansion, the computation of the coefficients $A^z, B^z$ involves truncations of series
and of the ranks of the matrices the determinants of which determine the coefficients.
Changing the order at which these are truncated would be a way to introduce error bars in
the MF results for $N<6$. Another would be to compute the coefficient ${\overline c_2}$ via local fits
as in \cite{MF12} as opposed to the global fits used here. 
This would require computing the Wilson Loops for increasing values of $L$ (starting from approximately $L=200$) and reading
off each time the plateau value for ${\overline c_2}$. The variations in the plateau values would introduce additional 
error bars on the MF data.
We did not do such an analysis here due to the lack of the necessary computing power.
Finally, clearly there will be new effects if we change the order of the truncation
of the MF expansion itself. Higher order effects will most certainly result in further corrections. 
It is hard to guess without further computations which of all possible corrections is mainly responsible for the discrepancy.

\section{Conclusion}

We developed the necessary formalism to perform the
mean-field expansion to first non-trivial order, for
five-dimensional, anisotropic $SU(N)$ lattice gauge theories in a
covariant gauge. We computed the mean-field background and then the
propagator, the scalar observable and the Wilson Loop. The mass of
the vector observable, being a geometric quantity, is the same as
for $SU(2)$. We then presented an application on the isotropic
lattice. We computed Coulomb's constant in five dimensions and
demonstrated that it becomes independent of $N$ in the large $N$
limit. At $N=\infty$ it converges to the value $k_5=0.0757$, which
amounts to approximately a $17\%$ deviation from the value predicted
by a holographic approach.

\end{document}